\documentstyle[12pt]{article}
\begin{document}
\input epsf
\renewcommand{\thefootnote}{\fnsymbol{footnote}}
\newcommand{\Od}{{\cal O}}
\newcommand{\lsim}   {\mathrel{\mathop{\kern 0pt \rlap
  {\raise.2ex\hbox{$<$}}}
  \lower.9ex\hbox{\kern-.190em $\sim$}}}
\newcommand{\gsim}   {\mathrel{\mathop{\kern 0pt \rlap
  {\raise.2ex\hbox{$>$}}}
  \lower.9ex\hbox{\kern-.190em $\sim$}}}
  \begin{center}
{\Large An introduction to the dark energy problem\footnote{Contribution to 
the proceedings of  "Space Astronomy: 
The UV window to the Universe", El Escorial, Spain, 2007 }} \\
\vskip 1.0cm
 {\large Antonio Dobado and  Antonio L. Maroto}
\vskip 0.5cm
 Departamento de
F\'{\i}sica Te\'orica,
 Universidad Complutense de
  Madrid, 28040 Madrid, Spain
\end{center}
\vskip 1.0cm

\begin{abstract}
In this work we review briefly the origin and history of the
cosmological constant and its recent reincarnation in the form of
the dark energy component of the universe. We also comment on the
fundamental problems associated to its existence and magnitude
which require and urgent solution for the sake of the internal
consistency of theoretical physics.
\end{abstract}


\section{Standard cosmology and Eintein's cosmological constant}
In 1915 Einstein introduced his famous field equations for gravity
in the context of General Relativity, namely:
\begin{equation}
R_{\mu\nu}-\frac{1}{2}Rg_{\mu\nu}=8\pi G T_{\mu\nu},
\end{equation}
where $R_{\mu\nu}$ is the Ricci tensor and $R$ is the scalar
curvature  corresponding to the  space-time metric $g_{\mu\nu}$ of
signature $(+,-,-,-)$, $G$ is the Newton constant and $T_{\mu\nu}$
is the matter energy-momentum tensor which for a perfect fluid has
the general form:
\begin{equation}
T_{\mu\nu}=(\rho + p)u_{\mu}u_{\nu}-pg_{\mu\nu},
\end{equation}
with $\rho$ and $p$ being the energy density and pressure
respectively and $u_\mu$ is the  four-velocity. From the fluid
proper frame this velocity is by definition  $(1,0,0,0)$ (we are
using natural units where $c=\hbar =1)$ and the energy-momentum
tensor reads $T_{\mu\nu}=diag(\rho,p,p,p)$.

Einstein field equations relate the space-time geometry (lhs) with
its energy content (rhs) and probably are one of the most
important landmarks in the whole history of physics. The covariant
derivative of the geometrical side identically vanishes thus
implying $D^\mu T_{\mu\nu}=0$ which are the equations of motion
which must be supplemented with the matter equation of state. The
Einstein equations describe properly at the classical level any
known gravitational phenomena. Moreover they can be used to study
the evolution of the universe as a whole. This can be done by
assuming the Cosmological Principle which establishes that, at the
largest scales, the universe is homogeneous and isotropic. In this
case we have the so called Friedmann-Robertson-Walker (FRW) metric
which can be written as:
\begin{equation}
ds^2=dt^2-a^2(t)\left(\frac{dr^2}{1- k r^2}+r^2(d\theta^2+\sin ^2
\theta)d\phi^2\right)
\end{equation}
where the $k$ parameter can take the values $1,0$ or $-1$. By
inserting this metric in the Einstein equations one gets two
independent equations for the cosmological scale function $a(t)$.
The Friedmann equation:
\begin{equation}
H^2\equiv\frac{\dot{a}^2}{a^2}=\frac{8 \pi G}{3}\rho-\frac{k}{a^2}
\label{Fried}
\end{equation}
where $H(t)=\dot{a(t)}/a(t)$ is the Hubble parameter and the
deceleration equation:
\begin{equation}
\frac{\ddot{a}}{a}=-\frac{4\pi G}{3}(\rho+ 3 p). \label{acc}
\end{equation}
The equation of motion (energy conservation) is just
$\dot{\rho}=-3H(p+\rho)$. The equation of state is typically
assumed to  have the general form $p = w \rho$. Thus we have for
example $w= 0$ for non relativistic matter (dust) or $w=1/3$ for
ultra-relativistic matter or radiation. Then the equation of
motion can be easily integrated to find $\rho\sim a^{-3(1+w)}$.
From the acceleration equation above it is clear that the sign of
$\rho_e=\rho + 3 p$ determines the acceleration state of the
expansion, i.e. for $\rho_e>0$ the expansion would be decelerated
whereas in the opposite case, the expansion rate would increase in
time.

Assuming an universe dominated by non relativistic matter ($p=0$)
we have three different possibilities for its evolution.
Introducing the critical density $\rho_c\equiv3 H_0^2/(8 \pi G$)
(from now on we use for any dynamical quantity $X(t)$ the notation
$X_0\equiv X(t_0)$, being $t_0$ the present time), we find: a
recollapsing closed universe with $k=1$ for $\rho_c < \rho$, an
open expanding universe with $k=-1$ for $\rho_c > \rho $, and the
limiting case which has $k=0$ and is also expanding, flat, and has
$\rho_e = \rho$. Notice that the expansion phases in all three
cases are decelerated.

Einstein did not like this result at all since it implies the
existence of a dynamical universe making not possible to have a
static universe as was his expectation and the accepted paradigm
for most people in those days. In fact, even before the above
results were found, Einstein decided in 1917 to modify his field
equations by introducing the cosmological constant term, since
from his original field equations it was clear that a closed and
static universe required to have $\rho_e=0$. The modified
equations read:
\begin{equation}
R_{\mu\nu}-\frac{1}{2}Rg_{\mu\nu}-\Lambda g_{\mu\nu}=8\pi G
T_{\mu\nu} \label{EinsteinLambda}
\end{equation}
where $\Lambda$ is the famous cosmological constant. Once this new
term is included in the Einstein equations many things change.
First of all Minkowski space is not a vacuum solution any more.
The Newtonian potential, valid for low energy fields and low
velocities, becomes:
\begin{equation}
V(r)=-\frac{G M}{r}-\frac{\Lambda}{6}r^2
\end{equation}
i.e. for $\Lambda >0$ the cosmological term produce some kind of
anti gravity or repulsive gravity. By using the modified equations
Einstein was able to find a closed and static solution with:
\begin{equation}
\Lambda=\frac{3}{a^2}-8 \pi G \rho. \label{tune}
\end{equation}
However this solution has two important drawbacks. First it
requires a fine tuning between density, cosmological constant and
the universe radius. Second, and probably more important, the
solution is not stable under small changes of the radius.

One useful and suggestive way to deal with the cosmological
constant term consists in moving it from the lhs to the rhs of the
Einstein field equations. This amounts to a redefinition of the
energy-momentum tensor:
\begin{equation}
T_{\mu\nu}\rightarrow \tilde T_{\mu\nu} = T_{\mu\nu} +
T_{\Lambda\mu\nu}
\end{equation}
where $T_{\Lambda\mu\nu}\equiv \Lambda g_{\mu\nu}/(8 \pi G)$, i.
e. the new density and pressure are $\tilde \rho = \rho +
\rho_\Lambda$ and $\tilde p = p + p_\Lambda$ with $\rho_\Lambda
=\Lambda /(8 \pi G)$ and $p_\Lambda=-\rho_\Lambda$. Therefore the
introduction of the cosmological constant is formally equivalent
to  the assumption of the existence of some kind of vacuum energy
present even when there is no matter at all. However this vacuum
energy (dark energy as denominated by M. Turner) has very strange
properties since it has $w_\Lambda = -1$. This means that for
positive $\Lambda$ it has negative pressure and its effective
density is $\rho_e=-2\rho_\Lambda$, i.e. it is also negative and
therefore produces accelerated expansion. In addition the
$\rho_\Lambda$ does not change during the evolution of the
universe as can be trivially obtained from the solution of the
equation of motion given above. It is possible to define the so
called cosmological parameter (which together with the value of
the Hubble parameter $H_0$ completely specify a cosmological
model), just by dividing the energy density of each component by
$\rho_c$ defined above, i.e. $\Omega_{M,R}=\rho_{M,R}/\rho_c$ for
matter or radiation, $\Omega_\Lambda=\rho_\Lambda/\rho_c$ and so
on. From the Friedmann equation above Eq.(\ref{Fried}), it is
possible to find that $\Omega_M+\Omega_\Lambda+\Omega_R
+\Omega_K=1$, where $\Omega_K=-k/(a_0^2H_0^2)$. All those facts
are quite important in order to understand the present state of
the universe as we will see in detail later.

Given the FRW metric it is not difficult to compute the physical
or proper distance. For rays coming from a distant object  at
constant $\theta$ and $\phi$ at  time $t$ we have:
\begin{equation}
d(t)=\int_0^r\sqrt{g_{00}}  dr'=
a(t)\int_0^r\frac{dr'}{\sqrt{1-kr'^2}}
\end{equation}
For a photon emitted from this object at time $t$ and arriving
now, i.e. at $t_0$, we have that the null geodesic equation
$ds^2=0$ implies that the wave length at origin $\lambda$ and the
received  wave length $\lambda_0$ are related by
$\lambda_0/\lambda=a(t)/a(t_0)$. It is customary  to define the
redshift parameter $z$ as $z\equiv (\lambda_0-\lambda)/\lambda$ or
$z=a_0/a-1$, i.e. it represents the fractional increase of
wavelength in such a way that for $z>0$ we have redshift and an
expanding universe and for $z<0$ we have blueshift and a
contracting universe. From the observational point of view it is
also interesting to introduce the so called luminosity distance.
If $L$ is the luminosity  of a distant object (total energy
emitted per unit of time), the flux $\Phi$ (energy received per
unit of area and time) is given by:
\begin{equation}
\Phi =\frac{L}{4\pi a_0^2r^2(1+z)^2}
\end{equation}
where the two $1+z$ factors come from the photon energy redshift
and the time dilatation between the emission and observation time
respectively. The luminosity distance $d_L$ is defined so that
$\Phi\equiv L/(4 \pi d_L^2)$ or in other words $d_L= a_0 r (1+z)$.
Now it is possible to expand this definition around $z=0$ for
$z<<1$ to find:
\begin{equation}
d_L=H_0^{-1}z\left(1+\frac{1-q_0}{2}z+...\right)
\end{equation}
where the deceleration parameter $q$ is defined as
$q=-\ddot{a}a/\dot{a}^2$. As commented above, for a matter
dominated universe we would have $q_0>0$. For small enough $z$ we
have:
\begin{equation}
z\simeq H_0 d_L.
\end{equation}
This is the famous Hubble law which was first found experimentally
in the twenties by Wirtz and Hubble by plotting the redshift
versus the luminosity distance of many galaxies around ours.
Hubble found a positive value for $H_0$ given by $H_0= 100 h$ km
s$^{-1}$ Mpc$^{-1}$
 where $h$ is some number of order one. As $H_0$ is
positive this shows that the universe is expanding  which is
probably one of the biggest scientific discoveries of all times.
One of the important theoretical consequences of this fact is
that, at least in principle, there is no need for introducing the
cosmological constant. It is sad to think that if were not for the
Einstein's prejudice about the static universe, he could have
predicted this expansion years before its discovery. It is usually
said that some time later Einstein declared that the introduction
of the cosmological constant was the biggest blunder he ever made
in his life.

\section{The accelerating universe and  dark energy}
In 1998 two independent teams, the Supernova Cosmology Project
(Perlmutter et al. (1999)) and the High-z Supernovae Search Team
(Riess et al.(1998)), extended the luminosity distance vs.
redshift relation to higher redshifts ($z\lsim 0.83$). For that
purpose they identified a new type of standard candle which was
sufficiently bright to be seen from very long distances, namely a
particular type of supernovae explosion known as Type Ia. This
type of supernovae are found in binary systems in which one of the
white dwarf stars exceeds the Chandrasekhar limits due to a
accretion from its companion. Although the absolute luminosity is
not the same for all SN Ia, it is found that the duration of the
explosion is related to the intrinsic luminosity and after
appropriate rescaling and corrrections a common light curve can be
found for all SN Ia.

Extending the Hubble diagram to high redshift allowed both teams
to measure the deceleration parameter and, unexpectedly, they
found a negative value for $q_0$, i.e. the expansion of the
universe would be accelerating today rather than decelarating as
one would expect in a matter dominated universe. A few years
later, the Hubble Space Telescope (Riess et al.(2004)), identified
16 new SNIa at $z>1.25$. The new data allowed to confirm not only
the present acceleration of the expansion rate, but also showed
the existence of an deceleration-acceleration transition at
redshift around $z_c\sim 0.5$. These results suggest a transition
around $z_c$ from a standard matter dominated universe to a
universe dominated by a new type of component with negative
pressure (responsible for the acceleration) as required by
Eq.(\ref{acc}).

The simplest explanation for that negative pressure fluid would be
the introduction of a cosmological constant, with equation of
state $w_\Lambda=-1$. Notice that in this case, the vaue of the
cosmological constant is not tuned as in Eq.(\ref{tune}) in order
to get and static Einstein universe, but instead its value would
be responsible for an accelerated expansion.

In such a case, when combining the data from SNIa with Cosmic
Microwave Background (CMB) anisotropies (Spergel et al.(2003)), it
was possible to determine the cosmological abundances of the
different components (see Fig. 1). Thus a good fit to the
cosmological observations would correspond to the so called
concordance model, i.e. a $\Lambda$CDM model with
$\Omega_M=0.27\pm 0.04$ and $\Omega_\Lambda=0.73\pm 0.04$. In
other words, the present universe would be mainly made of matter
and this new form of energy with negative pressure.

\begin{figure}[h]
\centerline{\epsfxsize=7.0 cm \epsfbox{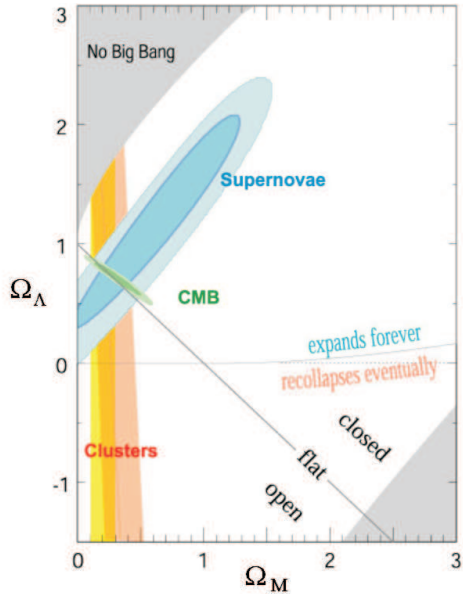}} \caption{Matter
density $\Omega_M$ vs. vacuum energy density $\Omega_\Lambda$
68$\%$ and 95$\%$ C.L. contours for supernovae, cluster and CMB
data, (from Supernova/Acceleration Probe: SNAP collaboration
(SNAP))}
\end{figure}

In the last years further observational evidence, independent of
SNIa observations have been obtained which supports the existence
of dark energy. In general, dark energy is believed to be a weakly
coupled component, its only effects being of gravitational nature
which makes its direct detection extremely difficult. So far the
observable effects of dark energy have been classified mainly in
two classes: modifications of the redshift-distance relation which
have been observed not only through SNIa, but also by means of the
so called baryon acoustic oscillations found in the large-scale
distribution of galaxies by the Sloan Digital Sky Survey  and 2dF
collaborations (Eisenstein et al. (2005),Cole et al. (2005)). On
the other hand, we have the effects on the growth of structure,
which could be measured through observations of galaxy clustering,
weak lensing or the integrated Sachs-Wolfe effect on CMB
anisotropies. These methods are currently in progress and new
surveys are planned to develop in the next decade (Trotta and
Bower (2006)).

\section{The cosmological constant problem and theoretical alternatives}
As we have seen, introducing a cosmological constant provides a
simple explanation for the current state of accelerated expansion
of the universe. However, a closer look at this solution reveals
some unpleasant features which we will discuss in this section.

The observational value $\Omega_\Lambda=0.73$ means that the
energy density of the cosmological constant is of the order of the
critical density, i.e. $\rho_\Lambda \simeq (10^{-3} \;{\mbox
eV})^4$. On the other hand the value of the Newton's constant is
$G\simeq (10^{19} \;{\mbox GeV})^{-2}$ in natural units. In other
words, if the cosmological constant is a true fundamental constant
of nature, the gravitational interaction described by Einstein
equations plus cosmological constant Eq.(\ref{EinsteinLambda})
would be controlled by two dimensional constants whose scales
$\sim 10^{19} \;{\mbox GeV}$ and $\sim 10^{-3} \;{\mbox eV}$ are
separated by more than 30 orders of magnitude. Explaining such an
enormous difference is called the cosmological constant problem.

\begin{figure}[h]
\centerline{\epsfxsize=7.0 cm \epsfbox{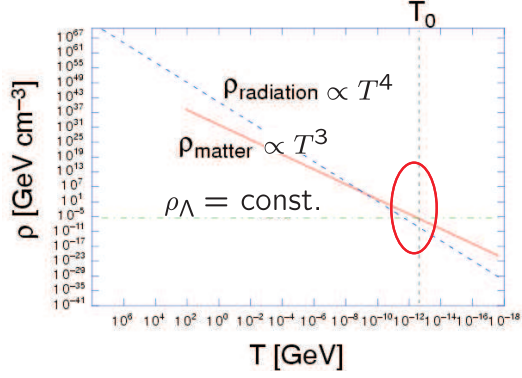}}
\caption{Evolution of the energy density of the different
components vs. photon temperature for radiation, matter and
cosmological constant. The circled area corresponds to the present
time.}
\end{figure}

A possible solution is that $\Lambda$ is not a true constant of
nature but some sort of effective term generated in the Einstein
equations by other physics. Thus for example, the vacuum energy of
all quantum fields present in the universe would contribute as
$\rho_\Lambda \sim\int_0^{M_*} d^3k k$, with $M_*$ the ultraviolet
cutoff of the theory. However the problem is not solved in this
way since we typically have $M_*\gg 10^{-3} \;{\mbox eV}$. There
are other candidates to dark energy apart from the cosmological
constant in the literature. Thus, it has been proposed  that dark
energy could be identified with the energy density of a dynamical
scalar field (quintessence) (Copeland et al (2006)). Such models
could exhibit equations of state today which deviate from $w=-1$
and could be discriminated from a cosmological constant by future
observations. However these models require appropriate potential
terms whose scales have to be fine tuned, in a similar way to the
value of the cosmological constant, in order for the accelerated
period to start at the correct time.

The dark energy problem can also be seen from a different
perspective. Since the time evolution of matter and  cosmological
constant are very different (see Fig.2), the fact that today they
have comparable values suggests that either we are living a sort
of cosmic coincidence, without deeper explanation, or there is a
strong relationship between the origin and evolution of the
different components of the universe. In any case, these problems
show why understanding the nature of dark energy has  become one
of the most important open questions in theoretical physics.

 {\bf Acknowledgements:}
 This work
 has been partially supported by the DGICYT (Spain) under the
 project numbers FPA 2004-02602 and FPA 2005-02327 and by the Universidad Complutense/CAM: project number
 910309. A. D. thanks Ana I. G\'omez de Castro for her kind
 invitation to participate in the NUVA Conference.


\begin{thebibliography}{}
\bibitem[Perlmutter et al. (1999)]{SCP} S. Perlmutter et al., {\it Astrophys. J.} {\bf 517}, 565 (1999)
\bibitem[Riess et al.(1998)]{Hz} A.G. Riess et al., {\it Astron. J.} {\bf 116}, 1009 (1998) and
{\bf 117}, 707 (1999)
\bibitem[Riess et al.(2004)]{Hubble} A.G. Riess et al., {\it Astrophys. J.} {\ bf 607}, 665 (2004)
\bibitem[Spergel et al.(2003)]{WMAP}
D. N. Spergel et al. \emph{Astrophys. J. Suppl.} \textbf{148},
175, (2003) and astro-ph/0603449.
\bibitem[SNAP]{SNAP} SNAP home page:
http://snap.lbl.gov/
\bibitem[Eisenstein et al. (2005)]{SDSS} D.J. Eisenstein et al., {\it Astrophys. J.} {\ bf 633}, 560 (2005)
\bibitem[Cole et al. (2005)]{2dF} S. Cole et al., MNRAS {\ bf 362}, 505 (2005)
\bibitem[Trotta and Bower (2006)]{Trotta} R. Trotta and R. Bower, {\it Astron. Geophys.} {\bf 47}:4:20-4:27,
 (2006)
\bibitem[Copeland et al (2006)]{quintessence} E.J. Copeland, M. Sami and S. Tsujikawa,
{\it Int. J. Mod. Phys.} {\bf D15} 1753, (2006).


\end{thebibliography}
\end{document}